\documentclass[12pt,preprint]{emulateapj}

\usepackage{epsfig}

\lefthead{Muzzin et al.}
\righthead{NIR Properties of Moderate-Redshift Galaxy Clusters}

\begin{document}

\title{Near-Infrared Properties of Moderate-Redshift Galaxy Clusters:
  Halo Occupation Number, Mass-to-Light Ratios and $\Omega_{m}$} 

\author{Adam Muzzin}
\affil{adam.muzzin@utoronto.ca \\ Dept. of Astronomy \& Astrophysics, University of Toronto \\
   50 St. George Street, Toronto, Ontario, Canada, M5S 3H4  } 
\author{H. K. C. Yee}
\affil{hyee@astro.utoronto.ca \\ Dept. of Astronomy \& Astrophysics, University of Toronto \\
   50 St. George Street, Toronto, Ontario, Canada, M5S 3H4} 
\author{Patrick B. Hall\altaffilmark{1}}
\affil{phall@yorku.ca \\ Department of Physics \& Astronomy,
York University \\ 4700 Keele Street, Toronto, Ontario, Canada, M3J 1P3} 
\author{H. Lin\altaffilmark{1}}
\affil{hlin@fnal.gov \\ Fermi National Accelerator Laboratory \\ P.O. Box 500, Batavia, IL 60510} 
\altaffiltext{1}{Visiting Astronomer, Kitt Peak National Observatory,
  National Optical Astronomy Observatory, which is operated by the
  Association of Universities for Research in Astronomy, Inc. (AURA)
  under cooperative agreement with the National Science Foundation.}

\begin{abstract}
 Using K-band imaging for 15 of the Canadian Network for Observational
 Cosmology (CNOC1) clusters we examine the near-infrared 
 properties of moderate-redshift (0.19 $< z <$ 0.55) galaxy clusters.
 We find that the number of
 K-band selected cluster galaxies within R$_{500}$ (the Halo Occupation
 Number, HON) is well-correlated with the the cluster dynamical
 mass (M$_{500}$) and X-ray Temperature (T$_{x}$); however, the
 intrinsic scatter in these scaling relations is 37\% and 46\%
 respectively.  Comparison with clusters in the local universe
 shows that the HON-M$_{500}$ relation does not evolve significantly between $z = 0$ and $z \sim 0.3$.  This suggests
 that if dark matter halos are disrupted or undergo significant tidal-stripping in
 high-density regions as seen in numerical simulations, the
 stellar mass within the halos is tightly bound, and not removed during the process.
 The total K-band cluster light (L$_{200,K}$) and K-band selected richness
 (parameterized by B$_{gc,K}$) are also correlated with both the
 cluster T$_{x}$ and M$_{200}$.  The total (intrinsic)
 scatter in the L$_{200,K}$-M$_{200}$ and
 B$_{gc,K}$-M$_{200}$ relations are 43\%(31\%) and 35\%(18\%)
 respectively and
 indicates that for massive clusters both L$_{200,K}$ and B$_{gc,K}$
 can predict M$_{200}$ with similar accuracy as T$_{x}$,
 L$_{x}$ or optical richness (B$_{gc}$). 
 Examination of the mass-to-light ratios of the clusters shows that
 similar to local clusters, 
 the K-band mass-to-light ratio is an increasing function of halo mass.    
 Using the K-band mass-to-light ratios of the clusters, we apply the
 Oort technique and find $\Omega_{m,0}$ = 0.22 $\pm$ 0.02, which agrees well with recent combined concordance
 cosmology parameters, but, similar to previous cluster studies, is on the
 low-density end of preferred values.     
\end{abstract}

\keywords{cosmology: dark matter $-$ large-scale structure of universe
\\ \hspace{15.0cm} galaxies: clusters: photometry $-$ fundamental parameters}

\section{Introduction}
One of the principal objectives in modern cosmology is a
good description of
the physics that govern the formation of structure
from the galactic through to the supercluster scale. 
Recent theoretical and observational work in this area has concentrated on understanding the relationship between the dynamically
dominant dark matter and the baryonic matter in the form
of stars and gas.  The major challenge has been that observables such as the
galaxy correlation function (e.g., Zehavi et al. 2005a, 2005b; Eisenstein et
al. 2005; Norberg et al. 2002, 2001), luminosity function
(LF; e.g., Babbedge et al. 2006; Ilbert et al. 2005; Dahlen et al. 2005;
Wolf et al. 2003; Blanton et al. 2003; Nakamura et al. 2003; Kochanek et al. 2001; Bell et al. 2003;
Cole et al. 2001), and the
Lyman-alpha forest (e.g., McDonald et al. 2006; Kim et al. 2002; Croft
et al. 2002) constrain the distribution of stellar 
mass and gas in the universe, yet dark matter is the dominant
gravitational component, and numerical simulations are more
effective at predicting its distribution than that of baryonic mass.. 
\newline\indent
Despite the difference between that which is easy to observe, and that
which is easy to simulate, recent combined N-body + Smoothed Particle
Hydrodynamics (SPH) simulations have suggested that dissipation from the baryonic
component is relatively unimportant in the process of galaxy formation
because the force of gravity from galactic dark matter halos is overwhelmingly
dominant (e.g., Weinberg et al. 2006; Nagai \& Kravtsov
2005).  These results naturally explain why purely N-body simulations
have been able to reproduce the correlation function (e.g., Col\'{i}n
et al. 1999; Berlind \& Weinberg 2002; Kravtsov et al. 2004; Tinker et
al. 2005; Nagai \& Kravtsov 2005; Tasitsiomi et al. 2004) and the number of satellite
galaxies per halo, the Halo Occupation Number (HON) or its
probability distribution, the Halo Occupation Distribution, (HOD,
e.g., Berlind \& Weinberg 2002, see Cooray \& Sheth 2002 for a review).  Furthermore, full SPH
simulations (e.g., Weinberg et al. 2006, 2004; Yoshikawa et al. 2001;
Pearce et al. 2001) and semi-analytic
models (e.g., Berlind et al. 2005, 2003; Cole et al. 2000; Zhu et al. 2006) which can be directly compared to observations, have
also begun to enjoy a
great deal of success at reproducing the LF,
correlation function, and HON.  These recent improvements in the quality
of simulations and semi-analytic models as well as new observational
datasets such as the SDSS and 2dFRS represent a major
breakthrough in our understanding of the relative distribution of
baryonic and dark matter in the universe and consequently, the
formation of large-scale structure.
\newline\indent
Detailed studies of galaxy clusters offer a complementary approach to
the correlation function, Lyman-alpha forest, and simulations for studying the relationship
between baryonic and dark matter.  Clusters allow us to probe
the distribution of baryons and dark matter in the most massive collapsed halos 
(which also happen to be the best-resolved objects in numerical
simulations).  The advantage of studying clusters is that
the baryonic content in the form of stars and
hot gas in the Intra-Cluster Medium (ICM) can be directly measured
with observations, and in addition to this,
cluster halos are sufficiently massive that their dark matter mass can be measured
using either weak lensing, X-ray data, or the dynamics of
cluster galaxies.  This allows direct comparison between the
baryonic and non-baryonic component on
a halo-by-halo basis, whereas for galaxy-mass halos, 
the mass-to-light ratio (M/L), HON, or bias is usually measured statistically using either galaxy-galaxy lensing
(e.g., Seljak et al. 2005;  Sheldon et al. 2004; Hoekstra et al. 2005, 2002), or by stacking
galaxies and studying the dynamics of satellite galaxies 
(e.g., Conroy et al. 2005, Brainerd et al. 2003), although recently
strong-lensing analysis of a sample of individual systems has been done (Treu et
al. 2006).  The advantage of
being able to study individual halos is that not only can the
correlation between baryonic and dark matter be measured (via the HON
or M/L ratio), but the relative {\it scatter} in the correlation can
also be determined.  Good constraints on the scatter are important because
it is a direct measure of the stochasticity in the baryonic/dark matter bias.
\newline\indent
In practice, the best way to measure the baryonic content of clusters in the
form of stars is with detailed modeling of the
stellar populations using either high-resolution spectroscopy, or
multi-band photometric observations.  As these observations are often
difficult to obtain for large samples of galaxies, K-band light has frequently
been adopted as a cheap and efficient proxy for
the stellar mass of galaxies.  K-band light is neither strongly affected by dust,
nor strongly enhanced by starbursts, and is therefore a good tracer of
underlying stellar mass of a galaxy (e.g., Brinchmann \& Ellis 2000;
Gavazzi et al. 1996; Rix \& Rieke 1993).  Furthermore, k-corrections
are generally small, and fairly independent of galaxy type (Poggianti 1997;
Mannucci et al. 2001).  Therefore, a study of the relative abundances of
K-band light and mass in
clusters is a good probe of the relationship between the cluster dark
matter mass and the baryonic mass in the form of stars.
\newline\indent
In addition to being useful for determining the relationship between
dark matter and baryonic matter,
a good understanding of the slope and scatter of the cluster K-band luminosity-mass
correlation and K-band richness-mass correlation will be
extremely valuable in the era of high-yield cluster surveys such as
the Red-sequence Cluster Survey 2 (RCS-2, Yee et al. 2007), the South Pole
Telescope (SPT, Ruhl et al. 2004), and the Atacama Cosmology Telescope
(ACT, Kosowsky 2006).  These projects
will use the evolution of the
cluster mass function, N(M, $z$) as a  probe of cosmological
parameters (see Mohr 2004 for a review); however, mass measurements for the thousands
of clusters that will be detected in these surveys are unlikely to be available from
traditional techniques such as L$_{x}$, T$_{x}$, weak-lensing, or
dynamics and therefore new, observationally cheaper proxies for cluster mass will be
required.  Two candidates for this mass indicator are the integrated
Sunyaev-Zeldovich Effect (SZE) $y$-parameter
(e.g., Motl et al. 2005), as well as the total cluster K-band luminosity
or K-band selected richness (e.g., L04, Ramella et al. 2004, Rines et
al. 2004).  For the SZE surveys the $y$-parameter has significant potential as a mass
indicator (e.g., Hallman et al. 2006, Motl
et al. 2005) as it is a measure of the mass of cluster baryons in the
form of hot gas.   For optical/IR surveys, K-band richness/luminosity is
an attractive alternative because it is a measure of the total mass
of baryons in the form of stars in cluster galaxies.  Gladders et al. (2006) have already shown that
optical richness works well as a cluster mass proxy, and 
using the optically-selected RCS-1 survey and an empirical
mass-richness calibration they determined cosmological parameters consistent with recent
concordance values (e.g., Spergel et al. 2006), thus illustrating the
potential for this technique.  K-band light is a
better tracer of stellar mass than optical light, and therefore it might be expected that the
scatter in the K-band light/richness vs. mass relation will be smaller than
the scatter in the optical light/richness vs. mass relation, or
at least might have fewer catastrophic outliers.
Using an observable with a smaller scatter, and fewer outliers could improve the accuracy in
the measured cluster mass function (e.g., Lima \& Hu 2005).
\newline\indent
Since the advent of the 2-Micron All Sky Survey (2MASS, Skrutskie et
al. 2006) several studies of the relationship between K-band light and
dark matter in local ($z <$ 0.1) clusters have been done.  Lin et
al. (2003, 2004; hereafter L04), Rines et
al. (2004), Kochanek et al. (2003), and Ramella et al. (2004) all show that the K-band selected
number counts and total cluster K-band light are indeed correlated with the
dark matter mass, and have rms scatters of $\sim$ 40\%.   L04, Rines et al. (2004), and
Ramella et al. (2004) also show that the slope of the L$_{K}$ vs. M$_{200}$ relation is shallower than
unity for clusters, suggesting that the conversion of baryons into stellar mass
proceeds less efficiently in higher-mass halos.  Recently, Lin et
al. (2006) presented an analysis of the K-band selected HON for a
heterogeneous sample of 27 clusters at 0 $< z <$ 0.9 and found no significant
evolution in the HON-Mass relation with redshift.  Interestingly, this
result was different
from their original analysis using a subsample of $\sim$ 20 clusters at 0.2 $<
z <$ 0.9 where they found the HON at fixed mass increases by a factor
of $\sim$ 2 at $z > 0.2$.  
\newline\indent
The purpose of this paper is to extend the analysis of the local
cluster K-band HON and
M/L ratio to higher redshift using a well-defined sample of clusters
with wide-field imaging and extensive spectroscopic data.
Our sample consists of 15 massive, X-ray selected clusters that were part of the
Canadian Network for Observational Cosmology (CNOC1, Yee et al. 1996)
project.  The K-band imaging, LF and density profiles for the clusters were presented in the
first paper in this series (Muzzin et al. 2006, hereafter Paper I).
In addition to the K-band imaging, these clusters also have
considerable optical spectroscopy and $g$ and $r$ photometry both of
which extend to R $\sim$ R$_{200}$ for each cluster. 
\newline\indent
The structure of this paper is as follows:  In $\S$2 we briefly
describe the dataset, and in $\S$3
present updated values for the cluster dynamical masses and X-ray
temperatures.  In $\S$4 we present the cluster HON and discuss its
redshift evolution.  Section 5 shows that total K-band luminosity
(L$_{200,K}$) and K-band selected richness (parameterized by B$_{gc,K}$)
are correlated with the cluster halo mass, and can potentially be used
as cheap proxies for this quantity in cluster abundance surveys.  In $\S6$ we present the K-band M/L ratios for the
CNOC1 clusters and in $\S$7 we present a measurement of
$\Omega_{m}$ using the Oort (1958) technique.  We conclude with a
summary in $\S8$.  When computing magnitudes and angular sizes we adopt an
$\Omega_{m}$ = 0.3, $\Omega_{\Lambda}$ = 0.7, H$_{0}$ = 70 km s$^{-1}$
  Mpc $^{-1}$ cosmology.
\section{Data}
The CNOC1 clusters are a set of 16 massive clusters with redshifts 0.17 $< z
<$ 0.54.  Fifteen of the clusters were
detected in the Einstein Medium Sensitivity Survey (EMSS, Gioia et
al. 1990).  Abell 2390 was added as the 16th cluster and is also a
massive cluster with strong X-ray emission.  The clusters have extensive optical photometry and
spectroscopy which were obtained as part of the CNOC1 project (Yee et
al. 1996).   
Our sample is comprised of 15 of the 16 CNOC1 clusters.  The
cluster MS0906+11 is omitted because it was  
shown to be a strong binary in redshift-space by Carlberg et al. (1996) and
therefore the mass measurement for the cluster is unreliable.  The
optical photometric and spectroscopic data as well as the K-band photometric data used in this paper were already presented in
Paper I.  We refer to that paper as well as the CNOC1 paper (Yee et
al. 1996) for complete details of the observations,
reductions and photometry and below present only a quick overview of the data.

\subsection{Optical Photometry and Spectroscopy}
Gunn {\it g} and {\it r} band imaging data were obtained at the 3.6m
Canada-France-Hawaii-Telescope (CFHT)
using the Multi-Object Spectrograph (MOS) camera.
Photometry was performed on these data using the Picture Processing Package (PPP, Yee
1991).  The photometry reaches a 5$\sigma$ depth of $\sim$ M$^{*}$ + 3 in both
the $g$ and $r$ bands.  The CNOC1
collaboration also obtained $>$ 2500 spectroscopic redshifts in the
fields of the 15 clusters using the CFHTMOS.   Of these, approximately one-half are 
cluster members.  The spectroscopy is sparsely sampled, but complete to a depth of K$^{*}$
+ 2 for all but the two highest-redshift clusters
(MS0016+16 and MS0451-03).  The
spectroscopy for those clusters is complete to $\sim$ K$^{*}$ + 1.

\subsection{Near Infrared Photometry}
K-band imaging for 14 of the 15 clusters was obtained at the
Kitt Peak National Observatory (KPNO) 2.1m telescope using the Ohio
State / NOAO Infrared Imaging Spectrograph (ONIS).  Observations were
taken in a Mauna Kea filter set version of the K$_{s}$ filter
(Tokunaga K),
which is nearly identical to the 2MASS K$_{s}$ filter.  For our
analysis we treat the Tokunaga K filter as the K$_{s}$-band and
hereafter refer to it as the ``K-band''.  K$_{s}$-band imaging of 
MS0440+02 was obtained using the PISCES camera on the Steward
Observatory 90$''$ telescope.  Photometry for the K-band data was also
performed using 
PPP.  The K-band imaging is complete to a 5$\sigma$ depth of $\sim$ K$^{*}$ + 2
for all clusters, except the two highest-redshift clusters where it is
complete to $\sim$ K$^{*}$ + 1 (see Table 1 of Paper I for a summary of observations).

\section{Cluster Physical Parameters}
The masses of the CNOC1 clusters are without a doubt the most
well-studied for clusters at moderate redshift.  There are numerous
measurements using X-ray temperatures (T$_{x}$, Hicks et al. 2006; Lewis et
al. 1999; Henry 2000; Mushotzky \& Scharf 1997), the
dynamics of cluster galaxies (van der Marel et al. 2000; Borgani et al. 1999; Carlberg et
al. 1996; 1997; Diaferio et al. 2005), and both strong (Fahlman et
al. 1994; Luppino \& Gioia 1995; Pierre et
al. 1996, Wu 2000) and weak lensing (Allen 1998; Hoekstra et al. 1998; Smail et al
1995; 1997).  In
this section we discuss the masses and dynamical radii we adopt for
comparing with the K-band properties.

\subsection{Dynamical Masses}
\indent
The first study of the dynamics of
the CNOC1 clusters was done by Carlberg et al. (1996) who  
measured the line-of-sight velocity dispersion ($\sigma_{1}$) for each
cluster.
These velocity dispersions were used to calculate the virial mass
(M$_{vir}$), the radius at which the cluster mass density
exceeds the critical density of the universe by a factor of 200
(R$_{200}$), and the mass contained within that radius (M$_{200}$), by assuming the cluster had an isotropic
velocity ellipsoid.  In subsequent work,
Carlberg et al. (1997) and van der Marel et al. (2000) examined in detail
the velocity dispersion profiles, and velocity anisotropy of the
clusters.  van der Marel et al. (2000) verified that they
are compatible with an isotropic velocity ellipsoid and that
the cluster mass profile was consistent with an NFW (Navarro
et al. 1997) profile.  The cluster velocity dispersions were also determined
by Borgani et al. (1999) using different background subtraction
techniques.  They found that the different techniques provide velocity dispersions that are
self-similar to $\sim$ 10\%, and that the velocity dispersions they derive are also
consistent to $\sim$ 10\% with the Carlberg et al. (1997) velocity
dispersions.  
\newline\indent
Unfortunately, the M$_{200}$ and R$_{200}$ values determined from these studies are now out
of date because they were computed assuming cosmological parameters
which are different from recent concordance values (e.g., Spergel
et al. 2006).  Fortunately, the velocity dispersions do not depend on
cosmology and, given that the clusters are consistent with an isotropic
velocity ellipsoid, we have recomputed both R$_{200}$ and M$_{200}$
with a $\Omega_{m}$ = 0.3,
$\Omega_{\Lambda}$ = 0.7, H$_{0}$ = 70 km s$^{-1}$ Mpc$^{-1}$
cosmology using the equations,
\begin{equation}
R_{200} = \frac{\sqrt{3}\sigma_{1}}{10 H(z)},   
\end{equation}  
and,
\begin{equation}
M_{200} = \frac{4}{3}\pi R_{200}^3 \cdot 200\rho_{c},    
\end{equation}
where H($z$) is
the Hubble constant at redshift $z$, and $\rho_{c}$ is the critical
density for a flat universe.
Equation 1 can be derived by assuming M$_{200}$ $\approx$ M$_{vir}$ and
equating the virial theorem (M$_{vir}$ = $\frac{3}{G}\sigma_{1}^{2}R_{vir}$) and Equation 2.
For all clusters the $\sigma_{1}$ values determined by Carlberg
et al. (1997) are used to compute M$_{200}$ and R$_{200}$. 
Errors in M$_{200}$ are calculated by propagating the errors in
$\sigma_{1}$ in the standard way.
The new values of R$_{200}$ and M$_{200}$ are on average $\sim$ 1.6 times larger
than the Carlberg et al. (1997) values.  We list the updated values in Table 1.
 
\subsection{X-ray Temperatures}
\indent
X-ray temperatures have been measured for the CNOC1 clusters by
several authors.  The clusters we originally discovered in the
Einstein Medium Sensitivity Survey (EMSS, Gioia et al. 1990); however,
only X-ray fluxes were computed using the original data.  Subsequently,  Lewis
et al. (1999) measured T$_{x}$ for 13 of the 15 clusters in
our sample using data from the ROSAT satellite.  They found that the
average cluster masses determined from T$_{x}$
 were in good agreement with the updated dynamical
masses, even though individual clusters could have discrepancies as
large as a factor of 2.
Recently, Hicks et al. (2006) used archival Chandra Advanced CCD
Imaging Spectrometer (ACIS) data to determine accurate
T$_{x}$ values for 13 of the 15 clusters in our sample.  They use
single and double-$\beta$-model fits to determine
the temperatures.  Several clusters (A2390, MS0839+29, MS1358+62,
MS1455+22) are also corrected for significant cooling
flows.  They find that the X-ray masses they
compute agree well with the dynamical masses determined with the
velocity dispersions of Carlberg et al. (1997) and Borgani et al. (1999).
These new T$_{x}$ values are superior to the older data and
for comparisons with the HON (\S 4), L$_{200,K}$ (\S 5.1), and B$_{gc,K}$ (\S 5.2)
we use these values.  There are two clusters 
without a Chandra observation (MS1224+20, MS1231+15).  For MS1224+20 we use the
temperature listed in Lewis et al. (1999).  MS1231+15 has no X-ray
temperature available so we assume a temperature of 6 keV, which is
appropriate given the X-ray luminosity (Yee \& Ellingson 2003).  We list the X-ray temperatures and their associated errors in Table 1.

\section{The Halo Occupation Number}
Clusters are massive dark matter halos, and therefore computing their HON
simply requires counting the total 
number of cluster members within some dynamical radius, typically R$_{200}$. 
The extensive spectroscopy for the CNOC1 clusters could
be used to differentiate between field and cluster galaxies; however, we prefer
to measure the HON using statistical background
subtraction.  Using spectroscopic
data to separate field and cluster galaxies is
preferable when computing cluster properties that depend on the luminosity of cluster
members (e.g., LFs,  total luminosity $\S$5.1, or M/L
ratios $\S$6.1) because these calculations require information on
the membership of individual galaxies to compute distance moduli and
k-corrections.  However, unless the spectroscopy of a cluster field is quite complete, statistical
background subtraction is better-suited for counting the
overdensity of cluster galaxies.  Furthermore, using statistical background
subtraction provides a consistent technique for each cluster and avoids
any biases from cluster-to-cluster that might be 
caused by poor determination of the spectroscopic selection function.
\newline\indent
The background counts are measured from our own K-band
imaging survey of CNOC2 fields (Yee et al. 2000).  The data was taken using
the PISCES camera on the Steward Observatory 90$''$ telescope and are reduced and
photometered using the same techniques as for the CNOC1 dataset.
The imaging consists of 23 (8$'$ $\times$ 8$'$) fields and covers a
total area of 0.26 square degrees.  Complete details of the
observations, data reduction, and photometry will be presented in a
future paper (Lin et al. in preparation).  Figure 1 shows background counts
from the CNOC2 fields with the number counts from the K20 Survey (Cimatti
et al. 2002) and the MUNICS Survey (Drory et al. 2001) overplotted
for comparison.  The CNOC2 galaxy
counts agree well with the counts from these surveys as well as various other
K-band surveys in the literature (e.g., Elston et al. 2006; Maihara et al. 2001; Saracco
et al. 2001; Gardner et al. 1993).
\newline\indent
We measure the HON for the clusters within the radius R$_{500}$ (the radius at which the cluster mass density
exceeds the critical density of the universe by a factor of 500), rather than the
more typical R$_{200}$.  It
would be preferable to use R$_{200}$, as it is similar to the cluster
virial radius; however, some of the clusters are
very sparsely sampled at R$_{200}$, and therefore require large
corrections to compensate for the poor sampling.  These corrections propagate
as large uncertainties in the total number of galaxies.  Conversely, the
coverage at R$_{500}$  is complete for all clusters (except MS1455+22) and
therefore those measurements have much smaller uncertainties and
should also be more robust.  The cluster R$_{500}$ is estimated from
the R$_{200}$ values assuming an NFW profile with c = 5.  Hereafter we refer to
the HON within R$_{500}$ as N$_{500}$ and the HON within R$_{200}$ as N$_{200}$.
\newline\indent
Thus far, the largest study of the cluster HON is the work of L04 who measured N$_{500}$
and N$_{200}$ for 93, X-ray selected, $z <$ 0.1 clusters using 2MASS data.  In
order to make a direct comparison between our sample and their sample, we
measure N$_{500}$ using the same limiting magnitude as L04.  Their
technique for measuring the HON involves determining the Schechter parameters K$^{*}$ and $\phi^{*}$ for each cluster
and then integrating the Schechter function
to an absolute magnitude limit of M$_{K}$ = -21 to find the total cluster counts.
For their ensemble of 93 clusters, they measured K$^{*}$ =
-24.02 $\pm$ 0.02.   Therefore, statistically speaking, their N$_{500}$ values are
measured at K$^{*}$ + 3, although the depth varies from
cluster-to-cluster.  Our data does not reach K$^{*}$ + 3 for most
clusters, 
therefore galaxies are counted to K$^{*}$ + 1 and then these counts
are extrapolated to K$^{*}$ + 3.
\newline\indent
K$^{*}$ evolves with redshift; however, we showed in Paper I that the evolution
tightly follows the luminosity evolution of a single-burst population
formed at high-redshift.  The single-burst model from Paper I is used to infer
K$^{*}$ at the redshift of each cluster.  The
background subtracted, K $<$ K$^{*}$ + 1 counts are scaled 
to the appropriate counts for K$^{*}$ + 3 by extrapolating a
Schechter function with a faint-end slope of $\alpha$ = -0.9.  This
value of $\alpha$ is consistent with the faint-end slope measured for
the CNOC1 cluster LF in Paper I ($\alpha$ = -0.84 $\pm$ 0.08) and is similar
to the faint-end slope measured for the L04 data ($\alpha$ = -0.84
$\pm$ 0.02).  
\newline\indent
We plot N$_{500}$ vs. M$_{500}$ for the clusters in the left panel of
Figure 2, and N$_{500}$ vs. T$_{x}$ in
right panel of Figure 2.  In both cases there is a good correlation
between the two parameters. The Spearman rank-correlation coefficients are 0.79 and
0.77 respectively, implying probabilities of 7.1 $\times$ 10$^{-4}$, 8.5
$\times$ 10$^{-4}$ that the data are uncorrelated.  
\newline\indent
One cluster which has an extremely small N$_{500}$ for
its mass is MS1455+22.  This cluster is also a significant outlier in
the M$_{200}$-B$_{gc}$ relation (B$_{gc}$ is a measure of cluster
richness, see $\S$5.2) 
from Yee \& Ellingson (2003, hereafter YE03).  They
suggest that the most logical interpretation is either that the mass has been
significantly overestimated or else that the cluster is in a very
advanced state of evolution, different from most clusters in this
redshift range. Therefore, MS1455+22 is excluded when we fit
both M$_{500}$ vs. N$_{500}$ and and T$_{x}$ vs. N$_{500}$ and
is plotted in Figure 2 as an open triangle.  Using a
$\chi^2$-minimization technique for errors in both parameters (Press et
al. 1992) we find the best-fit relation for the N$_{500}$ -
M$_{500}$ correlation is 
\begin{equation}
Log(M_{500}) = (1.40 \pm 0.22)Log(N_{500}) + (11.58 \pm 0.54),
\end{equation}
and has a reduced-$\chi^2$ of 1.60.  For the N$_{500}$ - T$_{x}$ the
best-fit relation is
\begin{equation}
Log(T_{x}) = (0.77 \pm 0.11)Log(N_{500}) + (-1.01 \pm 0.27),
\end{equation}
and has a reduced-$\chi^2$ of 2.46.  The correlations have Root-Mean
Squared (rms) scatters
of 51\% and 50\% respectively.  For the remainder of the
paper, the percentage rms scatter
in all correlations is computed using the same method as YE03.  The rms deviation from the fit in
terms of the dependent variable (in this case, M$_{500}$) is
determined and then compared to its mean value from the
entire sample.  
\newline\indent
The scatter in these correlations is larger than the
35\% scatter in the local scaling relations measured by L04.  However, the
random errors in our data are larger than theirs, and clearly a portion of the scatter comes
from these measurement errors.
Given that the total scatter is the result of both the
measurement errors in M$_{500}$ and N$_{500}$ as well as some
intrinsic scatter, we estimate the intrinsic scatter by
assuming that the measurement
scatter is independent of the intrinsic scatter and then subtracting the mean measurement
scatter of the dependent variable in quadrature from the total
scatter.  Using this simplistic method we estimate that the intrinsic scatter 
in N$_{500}$ - M$_{500}$ is 37\% and for N$_{500}$ - T$_{x}$ it is 46\%. 
\newline\indent
We compare the N$_{500}$ values from the CNOC1 clusters with the L04 values
in the top panel of Figure 3.  L04 measured N$_{500}$ by
integrating the cluster LFs over an NFW spatial distribution and
therefore it is the N$_{500}$ within a spherical volume.  Our
N$_{500}$ is measured using statistical background
subtraction and is therefore the number of galaxies in a cylinder of
radius R$_{500}$.  We convert our N$_{500}$ in cylinders to N$_{500}$
in spheres by multiplying by a deprojection constant of 0.791, the
value for the percentage difference in enclosed number of galaxies between
spheres and cylinders at R$_{500}$ for an NFW profile with c = 5.
The dash-dot blue line in Figure 3 is the best-fit relation for
the L04 clusters, and the solid red line is the best-fit relation for the
CNOC1 clusters.  The slope of the CNOC1 relations (M$_{500}$ $\propto$
N$_{500}^{0.71\pm0.11}$)
is slightly shallower than the L04 relation that includes the BCGs (M$_{200}$ $\propto$
N$_{500}^{0.82\pm0.04}$), but is consistent within the errors.  It is
also consistent with the scaling relation found by Rines et al. (2004)
for the CAIRNS clusters (M$_{200}$ $\propto$ N$_{200}^{0.70\pm0.09}$);
however, their relation is measured using N$_{200}$ and M$_{200}$ and a
different luminosity cut.  The
overall normalization of the CNOC1 N$_{500}$ is nearly identical to
the L04 normalization.  As a comparison between the samples
we plot the measured N$_{500}$ values
divided by those predicted from the local M$_{500}$-N$_{500}$
relation in the bottom panel of Figure 3 . Figure 3 shows that the HON - mass
correlation measured by L04 does not evolve between $z = $ 0 and $z
\sim$ 0.3.  The higher redshift CNOC1 clusters have  N$_{500}$ values
which are only 4\% $\pm$ 11\% smaller
than would be predicted by the local relation.  The mean(median) N$_{500}$/N$_{500,fit
  local} $ is 0.96(0.90) $\pm$ 0.11, and is consistent with no evolution in
the HON with redshift.  As a comparison, the L04
clusters have a mean(median) N$_{500}$/N$_{500,fit
  local} $ of 1.04(1.01) $\pm$ 0.04.
\newline\indent
The non-growth of the HON at a fixed mass with increasing redshift is
different from the results of L04.  In addition to their local
clusters, L04 also computed the HON for a subset of
clusters with $0.1 < z <$ 0.9 (K-band imaging from De Propris et al. 1999) which had available X-ray temperatures and
compared it to the local clusters.  Their comparison showed that
the HON was roughly a factor of 2 larger in the $z >$ 0.1 clusters.  From this
they suggested that half of all cluster galaxies may
be removed (via either merging or disruption) between $z \sim$ 0.6
(all but two of their clusters are $z <$ 0.6) and $z = 0$ in order to match the
local HON.  
\newline\indent
It is unclear why the conclusions of our  
studies are so different. We did note in Paper I
that it is challenging to reconcile the idea of a strong
evolution of the cluster HON between $z \sim$ 0.6 and $z = $ 0 and the passive
evolution of the K-band cluster LF.  Paper I showed
that the evolution of the cluster K$^{*}$ tightly follows the prediction of a
passive evolution model between $z \sim$ 0.5 and $z =
0$ and those results 
reaffirmed the findings of the pioneering cluster K-band luminosity
function study of de Propris et
al. (1999), as well as recent K-band LFs of
high-redshift clusters (e.g., Toft et al. 2003; Kodama \& Bower 2003;
Ellis \& Jones 2002; Strazzullo et al. 2006).  Interestingly, all of
the $z >$ 0.1 clusters used by L04 are part of the de Propris et
al. (1999) sample.  If 50\% of cluster galaxies are either merged, or
disrupted by tidal forces between $z \sim$ 0.6 and $z = 0$, then it
seems unlikely that the evolution of K$^{*}$ could follow a passive
evolution model, unless the probability of being merged or tidally
disrupted was independent of galaxy K-band luminosity (i.e., stellar mass).  While mass-independence may
be plausible in the case of
mergers, more massive galaxies are more difficult to disrupt.
If a significant number of disruptions do occur, they should
preferentially destroy galaxies on the faint-end of the luminosity
function and therefore will change its overall shape.  
\newline\indent
L04 appeal to the N-body simulations of Kravtsov et al. (2004)
for an explanation of the strong redshift evolution of the HON.  Kravtsov et
al. (2004) show that for dark matter halos, the HON at a given mass
becomes approximately a factor of three larger from $z =$ 0 to $z
=$ 5.  They show that the dark matter HON at a given mass is smaller at lower
redshift because low mass halos merge with high mass halos as well as
become tidally truncated or disrupted by massive halos in
high-density regions such as clusters.
\newline\indent
More recent simulations which incorporate both dark matter as well as
baryonic matter using Smoothed Particle
Hydrodynamics (SPH, e.g., Nagai \& Kravtsov 2005, Weinberg et
al. 2006) show that when the baryonic component is included, the
interpretation of the HON derived by counting galaxies becomes more subtle.  Nagai \& Kravtsov
(2005) simulated a set of 8 clusters and found that as subhalos enter the cluster R$_{200}$ they are stripped of
$\sim$30\% of their total dark matter mass.  Once they fall into the
cluster they are further stripped until $\sim$70\% of their original dark
matter mass is removed.  Their result confirms the earlier N-body
simulations which showed the dark matter HON is depleted with decreasing redshift,
especially in high-density environments; however, the SPH simulations also
show that primarily the dark matter is stripped from the halos and very little of the
baryonic mass (in the form of stars and gas) is lost
during this process.  The baryonic matter remains because it tends to
cool, and fall to the center of the subhalo and is amongst the
most tightly bound mass in the subhalo.
\newline\indent
These simulations suggest that the HON, when selected by counting the
number of massive dark matter subhalos is quickly depleted in high density regions such as clusters because of tidal
stripping and merging of the subhalos.  However, when the HON is selected by the
baryonic mass of halos
(i.e., using K-band number counts)
there should be little evolution in the HON at a given halo mass, because the
baryonic component of subhalos is tightly bound, and will only be
depleted in the case of full disruption.  This interpretation
is consistent with our data and is
also consistent with the observation that both K$^{*}$ (Paper I, L04,
de Propris et al. 1999) and $\alpha$ (L04) for cluster galaxies are basically identical
across more than an order-of magnitude in cluster mass (Paper I, L04,
de Propris et al. 1999).  Given that the
efficiency of tidal stripping is correlated with cluster mass, it is hard to
understand how the K-band LF of cluster galaxies
could be
independent of cluster mass unless the baryonic
component is unaffected in the process.
\newline\indent
In summary, we find that the K-band selected HON
of clusters at a given mass does not evolve significantly between $z =$ 0 and $z \sim$
0.3.  This result is consistent with the more recent work of Lin et al. (2006) who measured
the HON within R$_{2000}$, R$_{1000}$ and R$_{700}$ for a sample of 27
clusters 0 $< z <$ 0.9 (a few of which are 
included in L04) and also find no-evolution in the HON at fixed mass.
We argue that the non-evolution of the HON is naturally explained by recent SPH
simulations, which show that while dark matter halos encounter a  significant
amount of tidal stripping in high-density regions such as clusters, the
baryonic component within remains mostly intact. Furthermore, this
interpretation reconciles the strong tidal stripping with the purely passive evolution of the cluster
K-band LF, and its invariance across the cluster mass
spectrum.  
\newline\indent
Our data cannot rule out the possibility that the simulations may
over-exaggerate the tidal stripping and that similar to the baryonic
component, the dark matter 
mass-selected HON at a fixed cluster mass also does not evolve within the 
same redshift range.  A useful way to test this would be to compare
the M/L ratios of cluster galaxies and field galaxies at large radii
and look for evidence of truncated halos within the cluster
population.  Unfortunately,
this is a challenging task, because there are few tracers of the dark
matter potential at large radii for galaxies.
There is now evidence from galaxy-galaxy lensing studies of massive clusters that
supports the tidal-stripping scenario (Halkola et al. 2006, Natarajan
et al. 2002); however, in lower-mass 
clusters (M $\sim$ 10$^{13}$-10$^{14}$ M$_{\odot}$), the same trend is
not observed (Mandelbaum et al. 2006).  A more comprehensive study of
galaxy-galaxy lensing in clusters in the range of M $\sim$ 10$^{14}$-10$^{15}$ M$_{\odot}$ would
be particularly useful for testing this interpretation.
\begin{figure*}
\epsscale{1.0}
\plotone{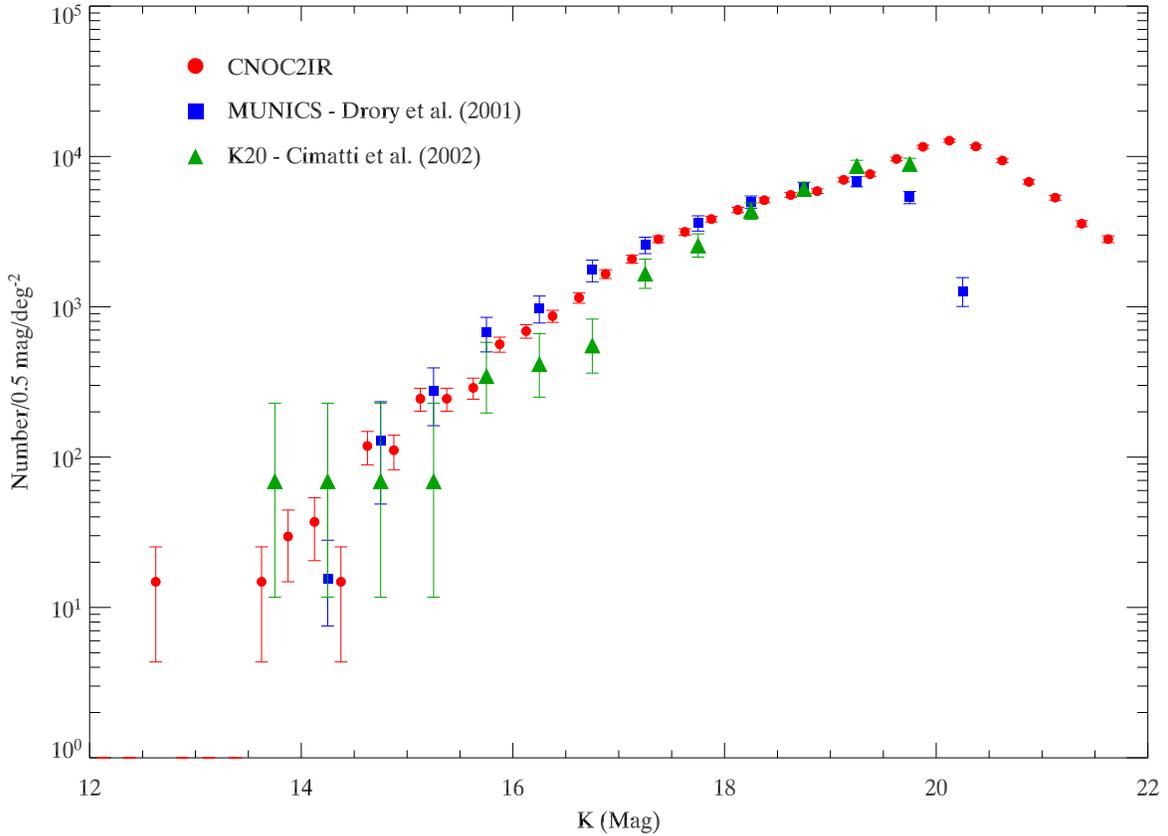}
\caption{\footnotesize {\it Red Dots}: Number of galaxies per square
  degree per 0.5 magnitudes as a function of magnitude determined from 0.26 square
  degrees of CNOC2 K-band imaging data.  {\it Green Triangles}: Number
  counts from the K20 Survey.  {\it Blue Squares}: Number counts from the
MUNICS Survey.}
\end{figure*}
\begin{figure*}
\plotone{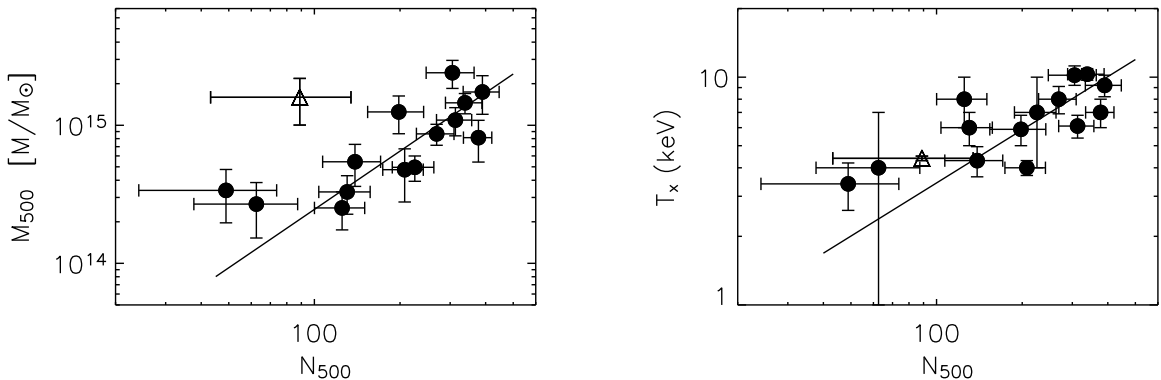}
\caption{\footnotesize Left Panel: Plot of  M$_{500}$ vs. N$_{500}$
  for the CNOC1 clusters.  The solid line is the
  best-fit linear relation and has an rms scatter of 50\%.  Right
  Panel: Plot of T$_{x}$ vs. N$_{500}$ for the same
  clusters.  The solid line is the best-fit linear relation and has an
rms scatter of 51\%.  MS1455+22 is plotted as an open triangle.}
\end{figure*}
\begin{figure*}
\epsscale{0.8}
\plotone{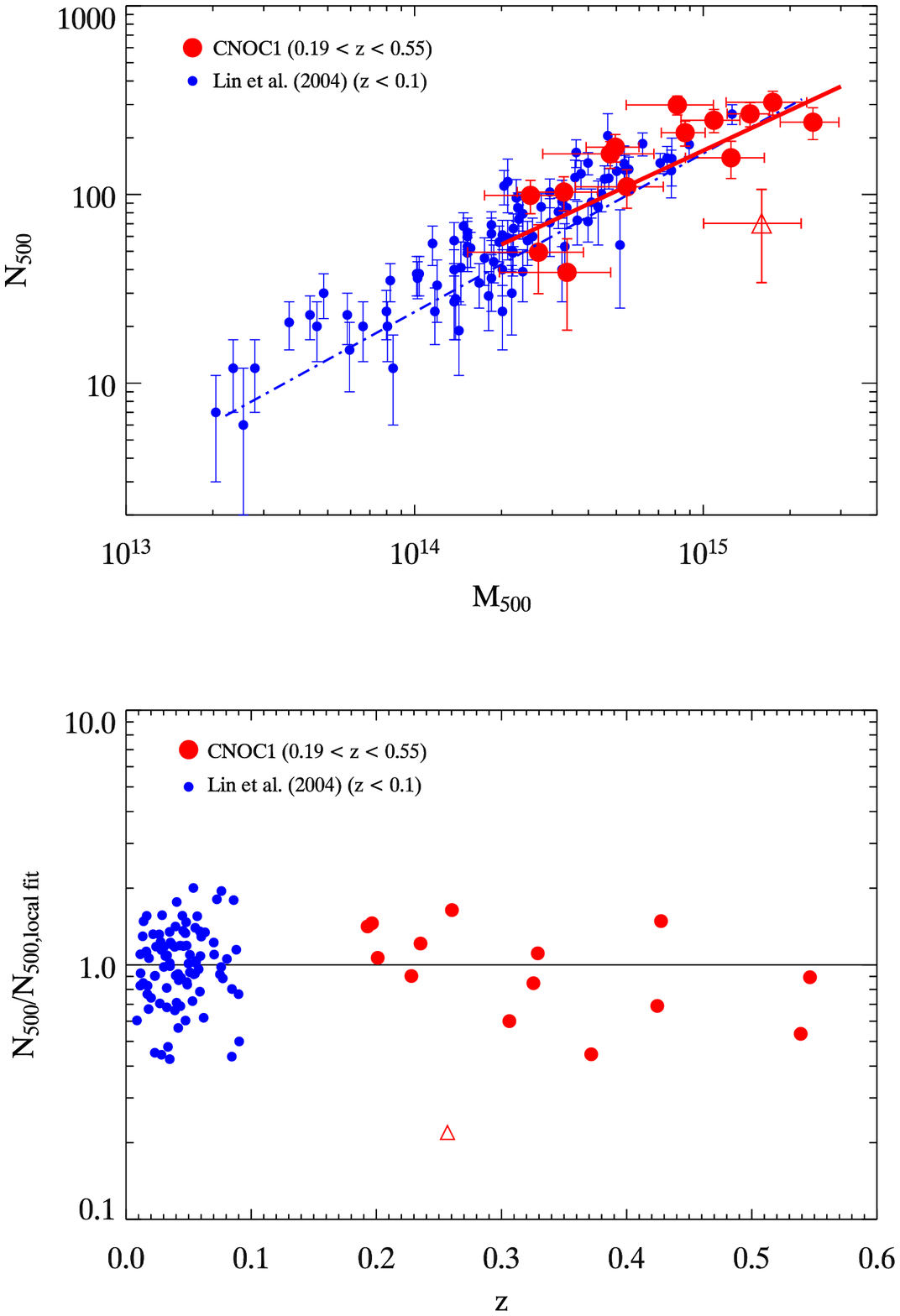}
\caption{\footnotesize {\it Top Panel}: Plot of N$_{500}$
  vs. M$_{500}$ for the CNOC1 clusters
  (large red points) and the L04 clusters (small blue points).
  MS1455+22 is plotted as an open triangle.  The
  solid red line is
  the best-fit relation for the CNOC1 clusters, and the dash-dot blue line is
  the best-fit relation for the L04 clusters.  The slope is the CNOC1
  correlation is slightly shallower, but consistent with the L04
  slope.  {\it Bottom Panel}: Plot of the measured N$_{500}$ divided
  by the value predicted by the L04 scaling relation as a function of
  redshift for the L04 clusters (blue points) and CNOC1 clusters (red
  points).  The CNOC1 normalization is consistent with the L04
  normalization suggesting no evolution in the HON with redshift.}
\end{figure*}
\section{K-Band Light and Richness as an Indicator of Cluster Mass}
In this section we measure the total K-band
luminosity within R$_{200}$ for the clusters (L$_{200,K}$) and examine its correlation with cluster mass.  We also
determine the correlation between K-band selected richness and cluster
mass.  
\subsection{Total K-Band Luminosity}
\indent
L$_{200,K}$ for a galaxy cluster is defined as the sum of the
K-band luminosity of all cluster galaxies within R$_{200}$ to a fixed
absolute magnitude.  
In principle
the measurement is straightforward; however, in practice, data are never
homogeneous and additional corrections must be applied.  In the case of the CNOC1
clusters, the clusters lie at a fairly large range of redshifts and therefore the
light from each galaxy must be both k-corrected and evolution
corrected to a common redshift.  The need for these corrections means
that the membership and approximate spectral-type for
each galaxy in the cluster field must be determined.  This is not
as precise using statistical
background subtraction and therefore we make use of the extensive
spectroscopic catalogues.  The 
spectroscopy is complete to K$^{*}$ + 1 for all clusters, but is sparsely sampled.
Therefore, a spectroscopic selection function is used to correct for the sampling.  The
method for determining the spectroscopic selection function
was developed by the CNOC1 (Yee et al.
 1996) \& CNOC2 (Yee et al. 2000) collaborations and is discussed in
detail in those papers as well as Paper I.  Similar to the analysis
in Paper I, we ignore the ``secondary'' selection effects (color, position, and $z$)
and use only the magnitude weights, which are the overwhelmingly
dominant selection bias (Yee et al. 1996).
\newline\indent
The k-correction for each galaxy is taken from the models of Poggianti (1997).
The k-corrections depend only mildly on spectral-type and the
color/spectral-type model
discussed in Paper I is used to estimate a spectral-type for each galaxy.  
Also, the LFs from Paper I 
showed that the luminosity evolution in the clusters is passive, but significant (0.35 $\pm$ 0.06 magnitudes from $z$ = 0
to $z$ $\sim$ 0.5).  Therefore, an
evolution correction from the Poggianti (1997) models is applied to each
galaxy, where we have transformed the Poggianti (1997) evolution corrections to our 
cosmology (see Paper I, \S 6.1.2).  By applying both k and evolutionary corrections 
the measured L$_{200,K}$ is corrected to $z$ = 0, and
therefore these values can be directly compared to local studies. 
\newline\indent
The L$_{200,K}$ for some clusters must also be corrected for the  
incomplete coverage of R$_{200}$.  We noted that for the HON (\S 4)
the coverage corrections 
propagate to large errors in the HON when it is computed within
R$_{200}$ and because of that, the HON was measured within
R$_{500}$.  The incomplete radial coverage is less problematic
when computing L$_{200,K}$ because much of the total cluster luminosity comes
from the BCG ($\sim$ 5 - 30\%).  Despite the fact that 
the  number of cluster galaxies at R$_{500}$ $<$ R $<$ R$_{200}$ is large, the
most luminous galaxies tend to reside in the cluster core, and
therefore galaxies at R$_{500}$ $<$ R $<$ R$_{200}$ 
contribute less to the total luminosity than to the HON.
\newline\indent
The clusters without complete coverage have been
observed in a strip through the cluster core; therefore, the coverage
problems are corrected by computing the
luminosity in circular shells.   The total luminosity
of a shell is multiplied by the ratio
of the total area of the shell, to the total area of the
shell with imaging/spectroscopic data.  
Several of the clusters (MS0016+16, MS0451-02, MS1006+12,
MS1008-12, and MS1455+22) have no coverage
for a few of the outermost shells.  For these clusters 
a correction based on the profile of clusters which have the best
coverage in the outer regions is applied.  This correction is small in four
of the five cases ($\sim$ 5-10\%), again because the majority of the cluster light comes
from the core.  MS1455+22 has a very large R$_{200}$ (which is likely
to be overestimated, see \S 4) and therefore the correction is much
larger ($\sim$ 60\%). 
\newline\indent
Once the selection function, k-correction and evolution corrections
have been applied, and   
the sum of the luminosity of galaxies brighter than K$^{*}$ + 1 has
been tabulated, the ``total'' luminosity of the cluster is finalized by extrapolating
the LF to a fixed absolute magnitude.  We choose to extrapolate to 
M$_{K}$ = -21 (i.e., K$^{*}$ + 3 at $z$ = 0), the same limiting absolute magnitude used by L04 and
and effectively the same limiting magnitude of the HON. 
We evolution-correct the combined LF from all 15
clusters to z = 0
(K$^{*}$ = -24.14 $\pm$ 0.15 and $\alpha$ = -0.84 $\pm$ 0.08, see Paper I) and use this for
extrapolation.  This LF is consistent with the average LF found by L04
(K$^{*}$ = -24.02 $\pm$ 0.02 and $\alpha$ = -0.84
$\pm$ 0.02) and because our clusters have been evolution corrected to
$z =$ 0, extrapolating the luminosity to M$_{K}$ = -21 results in a L$_{200,K}$ which
can be directly compared between the samples.
Before the correction is applied, the contribution from the BCGs is removed from the total cluster light.
The BCGs
contribute a significant, but variable, fraction of the total
cluster light ($\sim$ 5-30\% at K$^{*}$ + 1) and do not obey a Schechter
function.  Including them as part of
the extrapolation of the Schechter function would overestimate the
total cluster light.  The light from the BCG
galaxy is re-added to the total once the extrapolation has been done.   
Lastly, a bootstrap error for each cluster is computed by performing
the entire analysis using 300 resamplings with replacement from the data.
Table 2 lists the final L$_{200,K}$ values as well as the bootstrap errors.  
\newline\indent
We examine the correlation between the cluster mass and light by
plotting Log(M$_{200}$) vs. Log(L$_{200,K}$) in 
the first panel of Figure 4. The best-fit linear relation is 
\begin{equation}
Log(M_{200}) = (1.20 \pm 0.16)Log(L_{200,K}) - (0.95 \pm 2.21),
\end{equation}
and the fit has a reduced-$\chi^2$ of 2.65, where again we have
ignored MS1455+22 because it is likely to have an incorrect M$_{200}$.
The rms scatter in the
M$_{200}$ - L$_{200,K}$ relation is 43\% and from this we estimate
that the intrinsic scatter is 31\%.  The scatter in the 
correlation is about a factor of 1.3 larger than the 34\% total, and 24\% intrinsic scatter
measured by L04 for local clusters.  If we invert the axes, we find L$_{200,K}$
$\propto$ M$_{200}^{0.83\pm0.11}$.  This
agrees with the scaling relation for local clusters with the BCG
included which is L$_{200,K}$
$\propto$ M$_{200}^{0.72\pm0.04}$ (L04), and somewhat less well with the scaling relation for
local groups which is L$_{200,K}$ $\propto$ M$_{200}^{0.64\pm0.06}$ (Ramella et
al. 2004).  This shows that the slope, while slightly shallower for groups, does not
change significantly over $\sim$ 2 orders of magnitude in cluster mass.
Furthermore, the reasonable agreement between the slope and intrinsic scatter in the correlations at
different redshifts suggests that whatever physical processes
are responsible for building in the correlation have largely taken place by $z$ $\sim$ 0.3, and
that there is little change in the L$_{200}$ - M$_{200}$ relation from
$z \sim$ 0.3 to $z$
= 0, besides possibly a small decrease in the scatter.
\newline\indent
In the second panel of Figure 4 we plot the correlation between the cluster
X-ray temperature and L$_{200,K}$.  The best fit relation with
MS1455+22 excluded is
\begin{equation}
Log(T_{x}) = (0.77 \pm 0.08)Log(L_{200,K}) + (-9.45 \pm 1.07),
\end{equation}
and has a reduced-$\chi^2$ of 2.35.  The rms scatter in the
relation is 24\% and this corresponds to an intrinsic scatter of 14\%, notably better than the M$_{200}$ - L$_{200,K}$
relation.  As the CNOC1 clusters are all massive, relaxed X-ray
systems, this may not be surprising.  Using $r$-band data, YE03 found the correlation
between B$_{gc}$ and T$_{x}$ for the CNOC1 clusters to have a smaller scatter (21\%) than the
B$_{gc}$ - M$_{200}$ relation (31\%), although they used the
T$_{x}$ values determined by Lewis et al. (1999) rather than the
Hicks et al. (2006) values.
\newline\indent
If we compare the scatter
in the M$_{200}$ vs. L$_{200,K}$ relation to the scatter in the correlations of M$_{200}$ vs. T$_{x}$,
L$_{x}$, and B$_{gc}$ (optical) determined by YE03 for these clusters,
we find that the total K-band light is
approximately as accurate as those parameters in inferring the dynamical mass.  YE03 showed the
scatter in the M$_{200}$ vs. T$_{x}$,
L$_{x}$, and B$_{gc}$ (optical) for these clusters was 29\%, 35\%, and 31\%
respectively.  This demonstrates that L$_{200,K}$ is as good as, but no more
accurate than traditional mass indicators at inferring the cluster
dynamical mass.  It appears the fact that K-band light is a fairly clean tracer of an
individual galaxy's stellar mass does not result in a smaller scatter for the
L$_{200,K}$ - M$_{200}$ scaling relation, suggesting that the amount
of scatter measured in optical scaling relations is a real scatter in
relative amounts of stellar mass and dark matter mass in clusters, and not
enhanced due to a variance in star-formation properties or stellar populations
from cluster-to-cluster.
\newline\indent
It is possible that the scatter we measure is enhanced by systematics in the data
analysis caused by the need for k-corrections,
evolution corrections, spectroscopic selection functions, and coverage
corrections; however, the level of scatter we find
is of order that found by L04 in local clusters, which do not suffer from
those uncertainties.  Given that there is no reason to expect the
scatter in the scaling relations to become {\it smaller} at higher-redshift, it
would suggest that these corrections have been applied properly
and do not significantly increase the overall scatter.
\newline\indent
The correlations from Figure 4 are pleasing in the sense that
total K-band light is observationally much cheaper to measure than
L$_{x}$, T$_{x}$ or M$_{200}$ (via a velocity dispersion or weak lensing) for a given cluster.  For clusters in
this redshift range, a wide-field NIR camera on a 4m-class telescope can
achieve a limiting magnitude of M$_{K}$ $<$ K$^{*}$ + 1 with an
integration time of only a few minutes.  This means followup of many
candidates in the generation of high-yield cluster cosmology projects
is easily achievable.  Unfortunately, a major concern for cosmology projects is that
although L$_{200,K}$ appears to be a good
proxy for M$_{200}$ or T$_{x}$ it cannot be used in practice because
it requires an {\it a priori} knowledge of 
R$_{200}$.  In the
next section we show that the NIR-selected richness in a fixed
aperture is also well-correlated with the cluster mass and can be
used as a mass proxy.
\begin{figure*}
\plotone{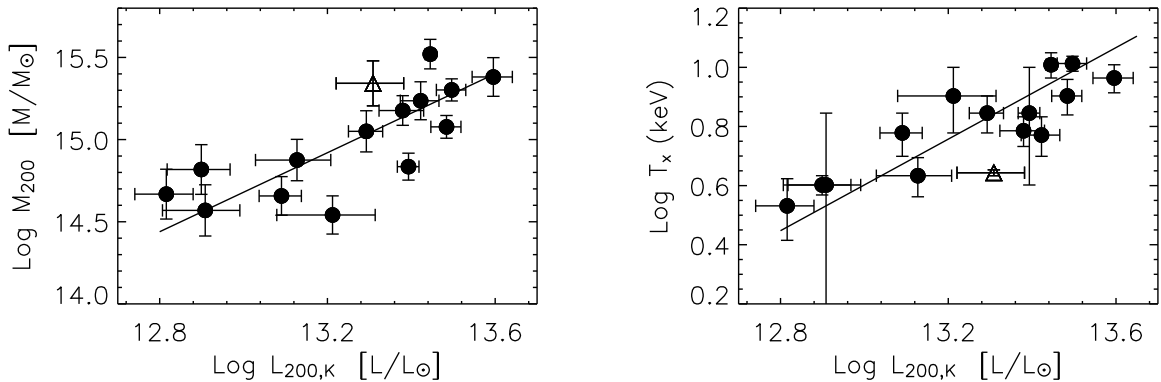}
\caption{\footnotesize Left Panel:  Plot of Log(M$_{200}$)
  vs. Log(L$_{200,K}$) for the CNOC1 clusters.  The solid line is the
  best-fit linear relation.  The rms scatter around the relation is
  43\% (31\% intrinsic).
  Right Panel:  Plot of T$_{x}$ vs. Log(L$_{200,K}$).  The solid line
  is the best-fit linear relation and as an rms scatter of 24\% (14\%
  intrinsic).  MS1455+22 is plotted as an open triangle in both panels
  and is excluded in the fits.
}
\end{figure*}
\subsection{B$_{gc}$ as an Indicator of Cluster Mass}
The good correlation between L$_{200,K}$ and M$_{200}$ found for the
CNOC1 clusters, and by other authors (e.g. Lin et al. 2003; L04; Rines
et al. 2004;
Ramella et al. 2004) demonstrates that K-band light can be used as an efficient 
estimator of the cluster halo mass.  Of course, measuring L$_{200,K}$
requires knowledge of R$_{200}$, which is generally an unknown, whereas the cluster
richness, parameterized by B$_{gc}$ does not require R$_{200}$.  B$_{gc}$ is the amplitude of the 3-dimensional,
spatial correlation function between the cluster center and the cluster
galaxies.  If the shape of the 2-dimensional angular
correlation function is assumed (i.e., w($\theta$) $\propto$
$\theta^{(1-\gamma)}$, with $\gamma$ $\sim$ 1.8), then its amplitude can be measured by counting
galaxies in a fixed aperture around the cluster center.  B$_{gc}$ is then
measured by deprojecting the angular correlation function via
($\xi(r) \propto r^{-\gamma}$) and scaling it by a luminosity
function.  A full derivation and motivation for B$_{gc}$ is presented in Longair
\& Seldner (1979).  The B$_{gc}$ parameter is defined as:
\begin{equation}
B_{gc} =
\mbox{N}_{net}\frac{(3-\gamma)\mbox{D}^{\gamma-3}\theta^{\gamma-1}}{2 \mbox{A$_{\theta}$}
  \mbox{I}_{\gamma}\Psi[\mbox{M}(\mbox{m}_{0},z)]},
\end{equation}
where N$_{net}$ is the background corrected cluster galaxy counts, D
is the angular diameter distance to the cluster redshift, $\theta$ is
the angular size of the counting aperture, A$_{\theta}$ is the angular
area of the counting aperture, I$_{\gamma}$
is geometric deprojection constant which depends on $\gamma$
(I$_{\gamma}$ = 3.78 for $\gamma$ $\sim$ 1.8), and $\Psi$ is the
integrated cluster LF up to the apparent magnitude M,
which corresponds to an absolute magnitude m$_{0}$ at the redshift
$z$.  The normalization of the LF ($\phi^{*}$) is the universal
normalization, and not the cluster normalization, so that B$_{gc}$ is
the overdensity of galaxies compared to the field density, not the
average cluster density.  This formula for
B$_{gc}$ is slightly different than formula
presented in Yee \& Lopez-Cruz (1999, their equation 3) in that it
contains the A$_{\theta}$ term.  We note that in their formula for
B$_{gc}$, Yee \& Lopez-Cruz (1999) had inadvertently left out the
$A_{\theta}$ term due to a transcribing error.
\newline\indent
At first, B$_{gc}$ may appear to be an unnecessarily complicated measure of the
cluster richness; however, using it has some
distinct advantages.  As interest in an observationally cheap proxy
for cluster mass has increased, so have the number of studies which
have looked at the correlation between cluster richness and mass
(e.g., Hansen et al. 2005; Miller et al. 2005; Gilbank et al. 2004;
L04; Rines et al. 2004; Ramella et al. 2004; Lin et al. 2003; Kochanek et al. 2003; YE03).  These studies define the
cluster richness as the number of galaxies within a fixed aperture, to
a fixed magnitude limit.  However, because different authors use different
apertures, and different magnitude limits, comparison between studies
is extremely difficult.  The advantage of the B$_{gc}$ parameter is that
it assumes a universal spatial distribution, and luminosity
function for clusters.  Therefore, aside from statistical fluctuations, {\it
  the value of B$_{gc}$ is, in principle, identical regardless of
what aperture is used to count galaxies, and which magnitude limit is
chosen}. 
\newline\indent
Yee \& Lopez-Cruz (1999) verified this was true,
but showed that measuring B$_{gc}$ from a fixed aperture with R = 500
kpc (using H$_{0}$ = 50 km s$^{-1}$ Mpc$^{-1}$)
produced the lowest statistical errors.  Furthermore, they showed that
counting galaxies to M $<$ M$_{*}$ + 1 was all that was required for
B$_{gc}$ to be robust.  Therefore, our values of B$_{gc,K}$ are
computed using a fixed
aperture of 500 kpc (although we use H$_{0}$ = 70 km s$^{-1}$ Mpc$^{-1}$) and by counting galaxies to K$^{*}$ + 1.
\newline\indent
Again, because B$_{gc}$ requires counting the number of cluster members, we use
statistical background subtraction, rather than the spectroscopic
weights.  YE03 demonstrated that B$_{gc}$'s computed using both
techniques agree extremely well.
In order to determine the integrated LF parameter
($\Psi$[M(m$_{0}$,$z$)]) a universal cluster luminosity
function must be assumed.  We showed in Paper I that besides the
passive evolution of the stellar populations, this is a reasonable
assumption for the CNOC1 clusters.  
When calculating $\Psi$[M(m$_{0}$,$z$)] we assume passive evolution of
the
LF of our ``average'' cluster 
(K$^{*}$ = -24.14, $\alpha$ = -0.84).  Unfortunately, we do not have a measurement of
the universal $\phi^{*}$ in units of Number/Mpc$^{3}$.  This cannot be determined from the cluster data alone because
clusters are high-density regions in the universe, and not
representative of the mean value of $\phi^{*}$.  Rather than using
values from studies of the K-band
field LF (which still have fairly large errors) we adopt the approach of Yee \& Lopez-Cruz
(1999) for determining $\phi^{*}$.  They
showed that a self-consistent value of $\phi^{*}$ could be estimated 
by evolving the cluster LF assuming a simple
model for the evolution of M$^{*}$({\it z}).  The normalization
($\phi^{*}$) is determined by requiring that the
integrated counts from the LF reproduce the
background galaxy counts.  We perform the same procedure
using a z$_{f}$ = 5.0 model passive evolution model (see Paper I) to
parameterize the evolution of M$^{*}$({\it z}).  By comparing the model background counts to the true background
counts and using a $\chi^2$-maximum-likelihood technique we determine $\phi^{*}$ =
4.34 $\times$ 10$^{-3}$ Mpc$^{-3}$.  
This value of $\phi^{*}$ is slightly larger than the $\phi^{*}$ = 3.40 $\pm$ 0.29
$\times$ 10$^{-3}$ Mpc$^{-3}$ determined
locally (Kochanek et al. 2001), and about 3 times as large as the $\phi^{*}$ = 1.78$^{+1.5}_{-0.9}$
$\times$ 10$^{-3}$ Mpc$^{-3}$ measured at 0.2 $<$ {\it z} $<$ 0.65 by
Pozzetti et al. (2003).  However, even if this value is incorrect it will be systematically
incorrect for all clusters and will only affect the intercept in the
correlation between
B$_{gc,K}$ and other physical parameters.  It will not affect the slope
or scatter, which are of principle interest.
\newline\indent
Column 6 of Table 2 lists the measured values of B$_{gc}$ for the
clusters.  The errors have been computed using the prescription from
Yee \& Lopez-Cruz (1999),
\begin{equation}
\frac{\Delta B_{gc,K}}{B_{gc,K}} = \frac{(N_{net} + 1.3^{2}N_{bg})^{1/2}}{N_{net}},
\end{equation}
where N$_{bg}$ is the number of background counts within the 500 kpc
and the 1.3$^2$ term is used to approximately account for the
clustering of background galaxies.
\newline\indent
In the left panel of Figure 5 we plot Log(B$_{gc,K}$) vs. Log(M$_{200}$)
for the clusters.  
The best fit linear relation with MS1455+22 excluded is 
\begin{equation}
Log(M_{200}) = (1.62 \pm 0.24)Log(B_{gc,K}) + (9.86 \pm 0.77).
\end{equation}
The fit has a reduced-$\chi^2$ of 1.29, and the rms scatter of the correlation is
35\% (18\% intrinsic).  This implies that the K-band selected richness is slightly
better than the total K-band light for predicting the mass of a galaxy
cluster.  The difference in total and (intrinsic) scatter between the two parameters is
notable, 35\%(18\%) in M$_{200}$ - B$_{gc,K}$
vs. 43\%(31\%) in M$_{200}$ - L$_{200,K}$ and this may be because of
the smaller aperture used to measure B$_{gc}$.   
\newline\indent
In the right panel of Figure 5 we plot Log(B$_{gc,K}$)
vs. Log(T$_{x}$).
The best fit linear relation with MS1455+22 excluded is 
\begin{equation}
Log(T_{x}) = (0.94 \pm 0.13)Log(B_{gc,K}) - (2.20 \pm 0.42),
\end{equation}
and has a reduced-$\chi^2$ of 2.00 and an rms scatter of 25\% (16\% intrinsic).  This
is very similar to the scatter seen in the L$_{200,K}$ - T$_{x}$
relation and shows that the fixed-aperture richness is an excellent
indicator of the cluster X-ray temperature at the $\sim$ 25\% level.
\newline\indent
If we compare the accuracy of B$_{gc,K}$ to the optical
B$_{gc}$ we find that they are almost identical at predicting
M$_{200}$ (scatters of 35\%, and 31\% respectively) as well as T$_{x}$
(scatters of 25\% and 21\% respectively).  B$_{gc,K}$ for the CNOC1
clusters also has a similar scatter to the fixed-aperture richnesses
measured for local clusters in the K-band.  L04 showed that the
scatter in the number of galaxies within 0.75 Mpc vs. M$_{500}$
was 43\%, and estimated that $\sim$ 24\% of the scatter was intrinsic.
The L04 cluster masses are measured using T$_{x}$ and therefore the
it is more relevant to
compare the L04 intrinsic scatter to the intrinsic scatter in the CNOC1 B$_{gc}$ - T$_{x}$
relation. The scatter in the L04 relation is somewhat larger than the
scatter 
from the CNOC1 relation, but this may be due
to the fact that the CNOC1 clusters are generally more massive than the L04
clusters.  
\begin{figure*}
\epsscale{0.9}
\plotone{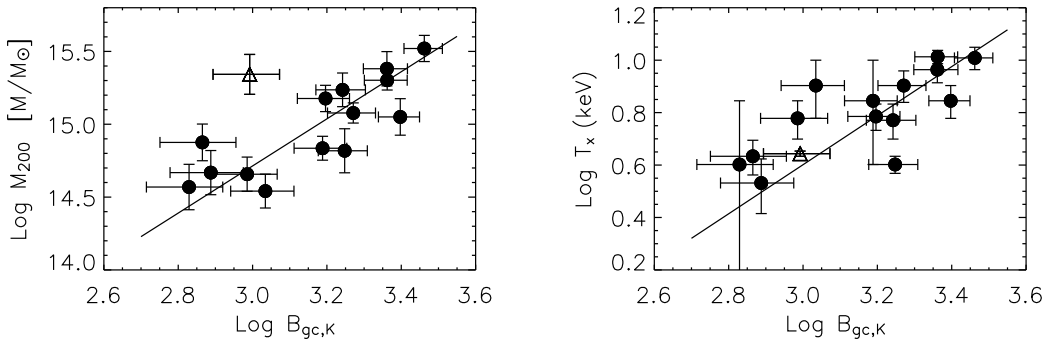}
\caption{\footnotesize Left Panel: Plot of  Log(M$_{200}$) vs. Log(B$_{gc,K}$)
  for the CNOC1 clusters.  The solid line is the
  best-fit linear relation and has an rms scatter of 35\% (18\% intrinsic).  Right
  Panel: Plot of Log(T$_{x}$) vs. Log(B$_{gc,K}$) for the same
  clusters.  The solid line is the best-fit linear relation and has an
rms scatter of 25\% (16\% intrinsic).  MS1455+22 is plotted as an open triangle in both panels
  and is excluded in the fits.}
\end{figure*}
\newline\indent
These results demonstrate that relatively
cheap IR imaging can be used to determine cluster masses with good
accuracy.  Specifically, IR imaging of even higher redshift clusters
($z >$ 0.5) will be valuable for determining masses in large optical
cluster surveys.  IR richnesses will likely be more robust
than optical richnesses at predicting cluster masses at
high-redshift because at high-redshift optical
bandpassess probe bluer parts of a galaxy's spectrum where the
star-formation properties  can drastically alter a galaxy's
luminosity.  
\newline\indent
It is also possible that in the case of large cluster surveys, B$_{gc,K}$
may prove to be as valuable for predicting the cluster halo mass as the
cluster velocity dispersion or X-ray temperature.
Determining the cluster mass from either the line-of-sight velocity
dispersion or X-ray temperature require assumptions about the
dynamical state of the 
cluster.  Dynamical masses require that the cluster is in virial
equilibrium and that that the shape of the velocity ellipsoid is known
(generally it is assumed to be isotropic).  X-ray masses require the
assumption of hydrostatic equilibrium, and spherical symmetry.  They
also require a correction if the cluster has a cooling flow.  In
cases where these assumptions do not apply (such as cluster-cluster
mergers, or a collapsing system in the process of forming) the masses of the clusters may be poorly estimated, and
these clusters will contaminate the measurement of the cluster mass function.  The number of catastrophic
outliers can have serious consequences on the measured cosmological parameters
because they are particularly sensitive to the number of rare,
massive clusters.  On the other hand, if
clusters which are unrelaxed, or currently undergoing a merger do not have 
the K-band light of their galaxies altered significantly, 
then using K-band richness as a mass indicator could potentially be a robust method for
determining the masses of clusters that are not in dynamical or
hydrostatic equilibrium.
\newline\indent
One way to test this hypothesis
is to use weak-lensing to measure the mass of clusters and
compare those masses to B$_{gc}$, and B$_{gc,K}$.
Weak-lensing does not require any assumptions about the dynamical
state of the clusters and therefore it is an unbiased (although
statistically noisy) measure of the
cluster mass.  Weak lensing masses for a subsample of the CNOC1
clusters have been measured and will be compared to B$_{gc,K}$ in a
future paper.

\section{The K band Mass-to-Light Ratio}
Recent studies of the K-band M/L ratio in local clusters (Lin et
al. 2003, L04, Rines et al. 2004, Ramalla et
al. 2004) have shown that the K-band cluster M/L
ratio is an increasing function of cluster mass (although Kochanek et
al. 2003 find it is roughly constant with mass).  
These studies have also found that the M/L ratio is a slowly decreasing
function of radius, with the integrated M/L ratio at R$_{200}$
(M$_{200}$/L$_{200,K}$) being
$\sim$ 15\% smaller than the M/L ratio at R$_{500}$ (e.g., L04).  
Here we present the first K-band M/L ratios of massive, intermediate-redshift
clusters.  
\newline\indent
We use the k and
evolution corrected L$_{200,K}$ values to compute
M$_{200}$/L$_{200,K}$ and therefore it is the M$_{200}$/L$_{200,K}$ ratio
of clusters corrected to $z$ = 0.  The values are listed in
Table 3 and the errors computed by propagating the M$_{200}$ and
L$_{200}$ errors in quadrature.
In Figure 6 we plot Log(M$_{200}$/L$_{200,K}$) vs Log(M$_{200}$).
There is a clear correlation of M/L with M$_{200}$ in the CNOC1 clusters.
The Spearman rank-correlation coefficient for these data is 0.60, which
implies a probability of 0.017 that the variables are uncorrelated.
The solid line in the plot is the best-fit linear relation with
MS1455+22 excluded:
\begin{equation}
Log(M_{200}/L_{200,K}) = (0.57 \pm 0.13) Log(M_{200}) - (6.92 \pm
2.04).
\end{equation}
The fit has a reduced-$\chi^2$ of 0.964 and implies that
M/L $\propto$ M$^{0.57 \pm 0.13}$.  
Interestingly, this slope is about a factor of 3 
steeper than the M/L $\propto$ M$^{0.17 \pm 0.11}$ that would be inferred using the Log(M$_{200}$)
- Log(L$_{200,K}$) relation (Eqn 5) where the variables are
less directly correlated (they still both depend on R$_{200}$).  As a
comparison, the inferred
relation is plotted as the dashed line in Figure 6.
The inferred relation, M/L $\propto$ M$^{0.17 \pm 0.11}$,
agrees well with the relation from local clusters where M/L $\propto$
M$^{\alpha}$ with $\alpha$ = 0.26 $\pm$ 0.04 (L04), and $\alpha$ = 0.31 $\pm$ 0.09
(Rines et al. 2004), but does not agree well with the value for local groups,
$\alpha$ = 0.56 $\pm$ 0.05 (Ramella et al. 2004).   The Ramella et
al. relation is similar to the 
correlation obtained by fitting Log(M/L) vs. Log(M$_{200}$) directly.
\newline\indent
It is likely that the inconsistent slopes from our own data, as well as between the
local cluster and group M/L vs. M$_{200}$ are primarily caused by the
large scatter in the M$_{200}$ - L$_{200,K}$ scaling relations and the
difficulties associated in fitting to data with a larger scatter than
is accounted for by the error bars.  It is probably  not caused by
real differences in the M/L vs. M$_{200}$ scaling relations for the
different cluster samples.  In our own data the M$_{200}$ -
L$_{200,K}$ relation has a large reduced-$\chi^2$ (2.65).  Examination
of Figure 4 suggests that the inflated
reduced-$\chi^2$ is not caused by a linear fit being
the incorrect model for these data, but is caused by a handful of significant outliers to
the relation.  
\newline\indent
Another consideration is that the intrinsic correlation
between M/L and M$_{200}$ (mass is in both parameters) can contribute
to the discrepancy in slope. 
The Ramella et al. (2004) masses and our own masses are determined from the
cluster velocity dispersion ($\sigma_{1}$) and equation 2, where M$_{200}$ $\propto$ $\sigma_{1}^{3}$.  The L04 masses are
determined from T$_{x}$ using the fit from Finoguenov et
al. (2001) where M$_{200}$ $\propto$ T$_{x}^{1.58}.$
Given that M$_{200}$ depends on $\sigma_{1}^{3}$ but only on
T$_{x}^{1.58}$, catastrophic errors in $\sigma_{1}$ will translate to
larger errors in M$_{200}$ than catastrophic errors in T$_{x}$ will. 
Consequently, given that the Log(M/L) vs. Log(M$_{200}$)
slope is $<$ 1, clusters which have their velocity dispersions
incorrectly measured (and not properly accounted for by the errors,
such as in the case of non-virialization) will steepen the slope of
the M/L vs. M$_{200}$ correlation because the mass term is in both
parameters.  
\newline\indent
As a test, we remove the highest and lowest mass CNOC1 clusters
(MS0440+02 and MS0451-03, both of which are significant outliers in
the M$_{200}$ - L$_{200}$ relation) and refit.
We find a slope of
$\alpha$ = 0.44 $\pm$ 0.15, which is more consistent with the
$\alpha$ = 0.17 $\pm$ 0.11 inferred directly from the M$_{200}$
vs. L$_{200,K}$ relation.
Unfortunately, this problem of outliers makes determining a robust
slope for the M/L vs. M$_{200}$ correlation difficult.
The data are less directly correlated in the case of 
Log(M$_{200}$) vs. Log(L$_{200,K}$), and therefore we return to that relation
and adopt M$_{200}$/L$_{200,K}$ $\propto$ M$^{0.17 \pm
  0.11}$ as the best slope for M/L vs. M$_{200}$ scaling relation for the CNOC1 clusters.  
\newline\indent
This slope, where M/L increases with increasing M$_{200}$ supports
the scenario proposed by both L04 and Rines et al. (2004), where
either the
star-formation efficiency of cluster galaxies is  a decreasing function of cluster mass,
or else that the amount of intra-cluster light is an increasing
function of cluster mass.  Interestingly, the slope of this scaling relation for the CNOC1
clusters is consistent with the the slope in local clusters and this shows that the M/L
vs. M$_{200}$ correlation
is in place by at least $z \sim$ 0.3, roughly 4 Gyr in lookback time.
\begin{figure*}
\epsscale{0.8}
\plotone{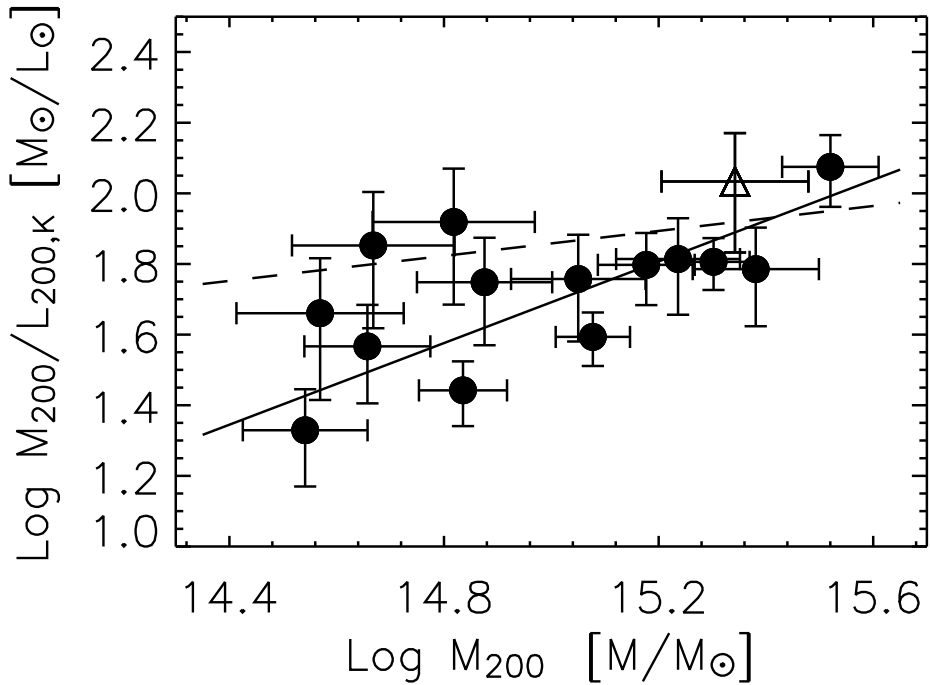}
\caption{\footnotesize Plot of Log(M$_{200}$/L$_{200,K}$) vs
  Log(M$_{200}$) for the CNOC1 clusters.  The solid line is the
  best-fit relation.  The dashed line is the relation that is inferred
from the Log(M$_{200}$) - Log(L$_{200,K}$) fit.  MS1455+22 is plotted as an open triangle
  and is excluded in the fit.}
\end{figure*}
\section{$\Omega_{m}$ from the Oort Technique}
The original purpose of the CNOC1 project was to measure the cosmological
density parameter $\Omega_{m,0}$ $\equiv$ $\rho_{o}$/$\rho_{c}$ using
the Oort (1958) technique.  The cluster M/L ratio, divided by the M/L ratio for closure (M/L)$_{c}$ $\equiv$
$\rho_{c}$/$j$, where $j$ is the field luminosity density, provides a
direct measure of $\Omega_{m,0}$ which is independent of the Hubble
parameter.   There has been some concern with this method because recent numerical
simulations suggest that light is a biased tracer of dark
matter, and that the M/L ratio of any region in the universe which
contains galaxies automatically provides an
underestimate of the universal M/L ratio, and an
underestimate of $\Omega_{m,0}$ (e.g., Ostriker et al. 2003).
However, this is only a significant concern for low-density regions in the universe (i.e.,
the group scale or less).  Unlike those lower density regions, rich clusters
are assembled from regions of $>$ 10 Mpc across
(Carlberg et al. 1997) and therefore, provided that observations
extend to the entire volume of the cluster (i.e., R $\sim$ R$_{200}$),
clusters should contain a sufficient collapsed volume to provide a representative
sample of the universal M/L ratio (e.g. Carlberg et al. 1996, Carlberg
et al. 1997).  This is yet to 
be verified in numerical simulations as, until recently, they lacked enough
volume to contain rich clusters.  The CNOC1 project was designed specifically
with this biasing concern in mind and therefore is a wide-field study
of a sample of rich clusters.
\newline\indent
Although biasing is not a major concern for the cluster technique, we know
that the stellar populations of galaxies in the cluster and field
environments are
different (e.g., Ellingson et al. 2001, Balogh et al. 1999, Poggianti et al. 1999, Dressler et
al. 1997) and without a good understanding on how these differences
affect the integrated luminosity of these regions, there arises the
potential for a systematic error in $\Omega_{m,0}$.
\newline\indent
Using the $r$-band data for the CNOC1 clusters, Carlberg et al. (1997) showed that $\Omega_{m,0}$ = 0.19 $\pm$
0.06 (random) $\pm$ 0.04 (systematic).  In their analysis they found that
the $r$-band cluster galaxies were on average 0.11 $\pm$ 0.05 mag
fainter than field galaxies at this redshift and hence they decreased
the cluster M/L ratio accordingly.
Given the need for this correction, it is advantageous to 
perform the same analysis with infrared data, because
infrared light is a better tracer of total stellar mass than
optical light.  The infrared luminosity of galaxies 
does not depend strongly on the current star-formation properties of
the stellar population and consequently, little or no
correction is required to account for the varying
star-formation properties of the field and cluster environment.  The only potential correction required
would be if the cluster environment played a role in the
star-formation {\it efficiency} of galaxies.  A differing
star-formation efficiency could manifest itself in two ways:
1) the stellar mass may be distributed
differently in the cluster/field environment (i.e., the {\it shape} of
the field and cluster LFs may be different); or 2) baryons may be converted
into stars at a different rate than in the field (i.e., the
ratio of the stellar mass to dark matter mass could be different
between these environments).  
\newline\indent
In terms of problem 1), Paper I showed that the
cluster K$^{*}$ was brighter than, but consistent with the field value ($\Delta$K$^{*}$ =
0.25 $\pm$ 0.26 mag).  Paper I also showed that 
$\alpha$ from the CNOC1 cluster LF was the same as $\alpha$
for local clusters as well as the local field.  Those results agreed well with local measurements
which showed the cluster K$^{*}$ and $\alpha$ are roughly the same as the
field (although we note that both L04 and Rines et al. 2004 find that
the cluster K$^{*}$ is $\sim$ 0.2 mag brighter at the 2$\sigma$
level).  Given that both K$^{*}$ and $\alpha$ for the cluster LF are
similar to the field LF, the distribution of galaxies in
terms of luminosity (and by corollary stellar mass) should be similar  in both
environments. 
\newline\indent
In terms of problem 2), it is important to note that the cluster M/L
ratio continues to be an increasing function of cluster mass, even for
objects as  massive as the CNOC1 clusters.  This shows that
the overall rate of converting baryons to stars is a function of
environment.  The rich CNOC1 clusters have a K-band M/L ratio which is about
an order of magnitude larger than the M/L ratio of groups (e.g., L04, Ramella
et al. 2004), and in turn, the group M/L ratio is about an order of
magnitude larger than for individual galaxies (e.g., Brinchmann \&
Ellis 2000).  Remarkably, despite the fact that rich clusters are two orders of magnitude less
efficient in converting baryons to stars than individual galaxies, their
continually increasing M/L ratio with increasing mass suggests
the possibility that they
may still be slightly biased tracers of stellar mass in the universe.  Unless
the M/L vs. M$_{200}$ relation flattens at the scale of $\sim$
10$^{15}$ M$_{\odot}$, the M/L ratio of clusters
could be lower than the universal value.  Unfortunately, there is no
data for the K-band M/L ratio of superclusters to determine if the M/L
ratio flattens at this size scale.
\newline\indent
The mean M/L ratio  for the clusters (corrected for passive
evolution to $z = 0$) is 61.2 $\pm$ 1.8 M$_{\odot}$/L$_{\odot}$.  Combining this with the
luminosity density of the local universe measured by Kochanek et
al. (2001) we find $\Omega_{m,0}$ = 0.22 $\pm$ 0.02
(random).  This result is in good agreement with combined constraints
from WMAP third-year results (WMAP3) and various other techniques (Spergel et
al. 2006).  The combined WMAP3 constraints on $\Omega_{m,0}$ range
from a low-density of $\Omega_{m,0}$ = 0.226$^{+0.030}_{-0.041}$ to a
high-density of $\Omega_{m,0}$ = 0.299$^{+0.019}_{-0.025}$.
Interestingly, the measurement from the clusters is closest to the lowest
density of the of preferred values from combined constraints.
Previous studies using K-band M/L ratios of local clusters have also
found values of $\Omega_{m,0}$ on the low-end of preferred values. 
\newline\indent
Both Rines et al. (2004) and Lin et al. (2003) measured $\Omega_{m,0}$
using their local cluster samples.  Rines et al. found that the result 
depended strongly on the location in the clusters where the M/L ratio was measured.  The cluster infall region had a
lower M/L ratio than the viralized region and therefore they found
$\Omega_{m,0}$ = 0.18 $\pm$ 0.04 from the virial region and
$\Omega_{m,0}$ = 0.13 $\pm$ 0.03 from the infall region.  Lin et
al. (2003) performed the same analysis with a larger sample of 27
clusters.  They found $\Omega_{m,0}$ = 0.17 $\pm$ 0.02 using the mean M/L ratio
of all clusters, and $\Omega_{m,0}$ = 0.19 $\pm$ 0.03 using a subsample
of their most massive clusters.
\newline\indent
Our analysis is in good agreement with both these studies, as well as
the $r$-band measurement from the same clusters by Carlberg et
al. (1997).  Our results further confirm that $\Omega_{m,0}$ from
clusters agrees well with concordance values but tends
to be on the low-density end of preferred values.  It is possible
that the cluster measurements prefer lower $\Omega_{m,0}$ because
their M/L ratios are still lower than the universal value.  A measurement of the K-band M/L ratio on the
supercluster scale would be a useful way to test this hypothesis.
Regardless, the fact that the cluster $\Omega_{m,0}$ still agrees well with
recent measurements of $\Omega_{m,0}$ using a variety of independent
techniques, and combinations thereof suggests that the cluster
M/L ratio is unlikely to be significantly lower than the universal value.
\section{Summary}
\indent
We have presented the K-band scaling relations for 15
moderate-redshift clusters with extensive optical spectroscopy and
wide-field K-band imaging.  The cluster HON is well-correlated with
M$_{500}$ and T$_{x}$; however, the intrinsic scatter in the scaling
relations at $z \sim$ 0.3 is fairly large (37\% and 46\%
respectively).  Comparing to local clusters we find that the HON is
consistent with no evolution at fixed cluster mass out to $z \sim$ 0.5.  This result, in tandem with the purely
passive evolution of the cluster LF, and the fact
that the cluster LF does not depend on mass suggests
that if the significant tidal stripping of galaxy halos in clusters
seen in numerical simulation occurs, the stellar mass within the
halo is tightly bound and remains intact.  This interpretation is
further supported by recent SPH simulations which show that the
baryonic matter is the amongst the most tightly bound mass within the halo. 
\newline\indent
Our data also show that both L$_{200,K}$ and B$_{gc,K}$ are well-correlated
with the cluster dynamical mass and X-ray temperature.  The slope of
the L$_{200}$ - M$_{200}$ relation at $z \sim$ 0.3 (L$_{200,K}$ $\propto$
M$_{200}^{0.83 \pm 0.11}$) is consistent with
the slope measured for local clusters, suggesting that the cluster
scaling relations are in place by at least $z \sim$ 0.3.  
The good correlation, and relatively small scatter (intrinsic +
measurement) in the B$_{gc,K}$ -
M$_{200}$ relation (35\%) and B$_{gc,K}$ - T$_{x}$ relation (25\%)
suggests that B$_{gc,K}$ would be useful mass indicator for upcoming cluster
cosmology projects.
\newline\indent
Our results from the M/L ratios of the clusters show that the
moderate-redshift CNOC1 clusters are very similar to local clusters in that
the M/L ratio is a slowly increasing function of cluster mass.  By comparing
the cluster M/L ratio to the local luminosity density we estimate that
$\Omega_{m,0}$ = 0.22 $\pm$ 0.02, which agrees well with the original
analysis of the CNOC1 clusters using optical data as well as with the
local estimates using K-band photometry.  The measured value is also
in good agreement with recent WMAP third year joint constraints;
however, similar to previous cluster studies the value is on the low-density end of preferred values. 
\newline\indent
The combined analysis of this paper and Paper I present a relatively
simple picture of the evolution of the near-infrared properties of 
clusters from $z = 0$ to $z \sim$ 0.3.  The correlation between the cluster near-infrared properties
(i.e., L$_{200,K}$, N$_{500}$, and M$_{200}$/L$_{200,K}$) and
M$_{200}$ shows no significant change between $z = 0$ and $z \sim$
0.3.  Furthermore, the scatter in these scaling relations is similar
at both redshifts.  The cluster LF shows only passive evolution
between $z = 0$ and $z \sim$ 0.3 and does not depend on cluster mass.
In addition to this, the (small) difference between the field and cluster LF
at $z = 0$ is unchanged at $z \sim$ 0.3.  These results all show that
there is little evolution in the the bulk of the stellar mass in
cluster galaxies over this redshift range besides the passive aging of
the stellar populations.  
\newline\indent
Specifically, it appears
that 1) the significant tidal
stripping of halo in high-density regions seen in N-body simulations does
not affect the stellar mass contained in galaxies nor change the
cluster scaling relations; 2) given passive
evolution of the LF, and no-evolution in the HON, mergers and
disruptions are
unlikely to play a significant role in cluster galaxy evolution at $z <$ 0.3;
and 3) that the changes seen in the cluster stellar populations over the redshift
interval $z = 0$ to $z \sim$ 0.3 (i.e., the increase in
blue-fraction/star-formation properties, and evolution of the
morphology-density relation) are ``superficial'' - they result from
changes in a small part of a galaxy's stellar mass, while the majority
of the stellar mass is already in place and passively evolving.
\newline\indent
Overall, it appears that the bulk of the stellar mass in cluster
galaxies is in place, and evolving passively with few mergers or disruptions between $z \sim$ 0.3 and $z = 0$, and that the
the cluster scaling relations are produced by processes that occur at higher redshifts. 
\acknowledgements
We thank the anonymous referee for a careful report which improved the
clarity of this paper.
A.M. would like to cite useful conversations and help from David Gilbank and
Kris Blindert which significantly improved the quality of the data
analysis.  A. M. acknowledges financial support from the National
Science and Engineering Research Council (NSERC) in the form of PGSA and PGSD2
scholarships.  The research of H.K.C.Y. is supported by grants from
Canada Research Chair Program, NSERC and the University of Toronto.
P.H. acknowledges support from NSERC.

\begin{center}
\begin{deluxetable}{lcccrrr}
\tablecolumns{7}
\tablecaption{Physical Properties of the CNOC1 Clusters}
\tablewidth{4.7in}
\tablehead{\colhead{Cluster} & \colhead{ $z$ } &
  \colhead{$\sigma_{1}$} & \colhead{R$_{200}$} & \colhead{ M$_{200}$} 
&
 \colhead{T$_{x}$} \\ 
\colhead{} & \colhead{} & \colhead{km s$^{-1}$} &
  \colhead{Mpc} & \colhead{M$_{\odot}$ $\times$ 10$^{14}$} & 
  \colhead{keV}  \\
\colhead{(1)}& \colhead{(2)}& \colhead{(3)}& \colhead{(4)}& \colhead{(5)}& \colhead{(6)}
}
\startdata
A2390       &  0.2279  & 1095 $\pm$ 61 & 2.00 &  20.1 $\pm$ 3.4 & 10.3$^{+0.6}_{-0.6}$ \\
MS0016+16   &  0.5466  & 1243 $\pm$ 128 & 1.89 & 24.0  $\pm$ 7.5 & 9.2$^{+1.0}_{-0.9}$ \\
MS0302+16   &  0.4246  & 656 $\pm$ 93 & 1.07 &  3.7  $\pm$ 1.6 & 4$^{+3}_{-1}$ \\
MS0440+02   &  0.1965  & 611 $\pm$ 62 & 1.13 &  3.5 $\pm$  1.1 & 8$^{+2}_{-1}$ \\
MS0451+02   &  0.2010  & 979 $\pm$ 76 & 1.81 &  15.0 $\pm$ 3.5 & 6.1$^{+0.7}_{-0.6}$ \\
MS0451-03   &  0.5392  & 1354 $\pm$ 105 & 2.06 & 33.3 $\pm$ 7.6 & 10.2$^{+1.0}_{-1.0}$ \\
MS0839+29   &  0.1928  & 788 $\pm$ 104 & 1.46 &  6.6 $\pm$ 2.7 & 4$^{+0.3}_{-0.3}$ \\
MS1006+12   &  0.2605  & 912 $\pm$ 101 & 1.63 &  11.2 $\pm$ 3.8 & 7$^{+1}_{-1}$ \\
MS1008-12   &  0.3062  & 1059 $\pm$ 107 & 1.85 & 17.2 $\pm$ 5.3 & 5.9$^{+0.9}_{-0.7}$ \\
MS1224+20   &  0.3255  & 798 $\pm$ 90 &  1.38 & 7.5 $\pm$ 2.5 & 4.3$^{+0.65}_{-0.65}$ \\
MS1231+15   &  0.2350  & 662 $\pm$ 69 & 1.20 &  4.5 $\pm$ 1.4 & 6$^{+1}_{-1}$ \\
MS1358+62   &  0.3290  & 910 $\pm$ 54 & 1.57 &  12.0 $\pm$ 2.1 & 8.0$^{+1.1}_{-0.9}$ \\
MS1455+22   &  0.2570  & 1169 $\pm$ 140 & 2.10 &  22.0 $\pm$ 8.2 & 4.4$^{+0.1}_{-0.1}$ \\
MS1512+36   &  0.3726  & 697 $\pm$ 96 & 1.17 & 4.7 $\pm$  2.0 & 3.4$^{+0.8}_{-0.7}$ \\
MS1621+26   &  0.4274  & 833 $\pm$ 55 & 1.36 & 6.8 $\pm$ 1.4 & 7$^{+3}_{-2}$ \\

\enddata
\end{deluxetable}
\end{center}
\begin{deluxetable}{lccccrrrr}
\tablecolumns{9}
\tablecaption{NIR Properties of the CNOC1 Clusters}
\tablewidth{7.2in}
\tablehead{\colhead{Cluster} & \colhead{L$_{200,K}$} &
  \colhead{$\epsilon$ L$_{200,K}$} 
& \colhead{(M/L)$_{200,K}$} & \colhead{$\epsilon$ (M/L)$_{200,K}$} &
  \colhead{B$_{gc,K}$} & \colhead{$\epsilon$ B$_{gc,K}$} & \colhead{N$_{500}$} & \colhead{$\epsilon$ N$_{500}$}\\ 
\colhead{} & \colhead{L$_{\odot} \times$ 10$^{13}$} &
  \colhead{L$_{\odot} \times$ 10$^{13}$} & \colhead{} & \colhead{} &
  \colhead{Mpc$^{1.8}$} &
  \colhead{Mpc$^{1.8}$} & \colhead{} & \colhead{}   \\
\colhead{(1)}& \colhead{(2)}& \colhead{(3)}& \colhead{(4)}&
\colhead{(5)}& \colhead{(6)}& \colhead{(7)}
& \colhead{(8)} & \colhead{(9)}
}
\startdata
A2390       &  3.14  & 0.25 & 64 &  14 & 2302 & 303 & 338 & 50 \\
MS0016+16   &  3.94  & 0.43 & 61 &  24 & 2298 & 311 & 389 & 56 \\
MS0302+16   &  0.81  & 0.17 & 45 &  25 & 674 & 157 & 63 & 25 \\
MS0440+02   &  1.63  & 0.43 & 21  & 8  & 1083 & 210 & 125 & 25 \\
MS0451+02   &  2.40  & 0.29 & 63 & 18 & 1571 & 252 & 313 & 44 \\
MS0451-03   &  2.79  & 0.08 & 119 & 34 & 2897 & 343 & 305 & 59 \\
MS0839+29   &  0.79  & 0.14 & 83 & 43 & 1768 & 266 & 207 & 34 \\
MS1006+12   &  1.96  & 0.18 & 57 & 24 & 2496 & 316 & 377 & 43 \\
MS1008-12   &  2.64  & 0.28 & 65 & 28 & 1745 & 266 & 198 & 44 \\
MS1224+20   &  1.34  & 0.27 & 56 & 24  &  733 & 169 & 139 & 32 \\
MS1231+15   &  1.23  & 0.14 & 37 & 14 & 967 & 198 & 130 & 27 \\
MS1358+62   &  3.05  & 0.25 & 39 & 8 & 1866 & 275 & 269 & 41 \\
MS1455+22   &  2.03  & 0.37 & 108 & 50 & 982 & 199 & 89 & 46 \\
MS1512+36   &  0.65  & 0.10 & 71 & 37 &  772 & 172 & 49 & 25 \\
MS1621+26   &  2.47  & 0.15 & 27 & 7 &  1543 & 249 & 225 & 38 \\

\enddata
\end{deluxetable}

\end{document}